\documentclass[11pt,a4paper]{article} %UKenglish,
\usepackage{enumerate}
\usepackage{textcomp}
\usepackage{url}
\usepackage{graphicx}
\usepackage[numbers]{natbib}
\usepackage{bussproofs}
\usepackage{tikz}
\usepackage{amssymb}
\usepackage{amsmath}
\usepackage{amsthm}
\usepackage{verbatim}
\usepackage[all,cmtip]{xy}
\usetikzlibrary{positioning, automata}
\usetikzlibrary{decorations.pathmorphing}
 \usetikzlibrary{snakes}

\tikzset{snake it/.style={decorate, decoration=snake}}
\newtheorem{lemma}{Lemma}
\newtheorem{theorem}{Theorem}
\newtheorem{definition}{Definition}
\newtheorem{example}{Example}
\newtheorem{corollary}{Corollary}

\begin{document}

\sloppy

\title{Rational index of bounded-oscillation languages\footnote{%
This research was supported by the Russian Science Foundation, project 18-11-00100.}}

\author{Ekaterina Shemetova\footnote{%
Department of Mathematics and Computer Science, St. Petersburg State University, 
7/9 Universitetskaya nab., Saint Petersburg 199034, Russia.}
\footnote{%
St. Petersburg Academic University, 
ul. Khlopina, 8, Saint Petersburg 194021, Russia.}
\footnote{%
JetBrains Research,
Primorskiy prospekt 68-70, Building 1, St. Petersburg, 197374, Russia.}
\and
Alexander Okhotin\footnotemark[2]
\and
Semyon Grigorev\footnotemark[2] \footnotemark[4]
}

\maketitle

\begin{abstract}
The rational index of a context-free language  $L$ is a function $f(n)$, such that for each regular language $R$ recognized by an automaton with $n$ states, the intersection of $L$ and $R$ is either empty or contains a word shorter than $f(n)$. It is known that the context-free language (CFL-)reachability problem and Datalog query evaluation for context-free languages (queries) with the polynomial rational index is in NC, while these problems is P-complete in the general case. We investigate the rational index of bounded-oscillation languages and show that it is of polynomial order. We obtain upper bounds on the values of the rational index for general bounded-oscillation languages and for some of its previously studied subclasses.

\end{abstract}

\section{Introduction}
\label{intro}
The notion of a rational index was introduced by Boasson et al. \cite{RatBasic} as a complexity measure for context-free languages.  The rational index $\rho_L(n)$ is a function, which denotes the maximum length of the shortest word in $L \cap R$, for arbitrary $R$ recognized by an $n$-state automaton. The rational index plays an important role in determining the parallel complexity of such practical problems as the context-free language (CFL-)reachability problem and Datalog chain query evaluation.

The CFL-reachability problem for a fixed context-free grammar $G$ is stated as follows: given a directed edge-labeled graph $D$ and a pair of nodes  $u$ and $v$, determine whether there is a path from $u$ to $v$ labeled with a string in $L(G)$.  That is, CFL-reachability is a kind of graph reachability problem with path constraints given by context-free languages. It is an important problem underlying some fundamental static code analysis like data flow analysis and program slicing \cite{RepsBasic}, alias analysis \cite*{Chatterjee, alias}, points-to analysis \cite{Incremental} and other \cite{Cai, android, typeflow}, and graph database query evaluation \cite{Azimov, GrigorevRagozina, HellingsCFPQ, RDF}.

The Datalog chain query evaluation on a database graph is equivalent to the CFL-reachability problem. 
\begin{example}[Datalog query as a context-free grammar]
\label{DatalogExample}
Consider a database  $D$ with relation ``$child$''. It also can be represented as a digraph $G$, where each node of the graph corresponds to a person, and edges are labeled with a word ``$child$''.

The following Datalog query determines all pairs of people $x$ and $y$ such that $x$ is a descendant of $y$:

$Desc(x, y)$ :- $Child(x, y)$

$Desc(x, y)$ :- $Child(x, z), Desc(z, y)$

This query can be represented as a context-free grammar with the following rules:

$Desc \rightarrow Child$ $\vert$ $Child$ $Desc$

$Child  \rightarrow child$

Thus, evaluating the above mentioned Datalog query over database $D$ is equivalent to solving the CFL-reachability problem for a context-free grammar representation of this query and the edge-labeled graph $G$.
\end{example}

Unlike context-free language recognition, which is in NC (when context-free grammar is fixed), the CFL-reachability problem is P-complete \cite{ PCompl, RepSeq,  Yannakakis}. Practically, it means that there is no efficient parallel algorithm for solving this problem (unless P $\neq$ NC).

The question on the parallel complexity of Datalog chain queries was investigated independently \cite{ChainQ, Vardi, Ullman}. Ullman and Van Gelder \cite{Ullman} introduce the notion of a \textit{polynomial fringe property} and show that chain queries having this property is in NC. The polynomial fringe property is equivalent to having the polynomial rational index: for a context-free language $L(G)$ having the polynomial rational index $\rho_L(n) = poly(n)$, where $poly(n)$ is some polynomial, is the same as for corresponding query to have the polynomial fringe property. It has been shown that for every algebraic number $\gamma$, a language with the rational index in $\Theta (n^\gamma )$ exists \cite{GreibRat}.  In contrast, the rational index of languages, which generate all context-free languages (an example of such language is the Dyck language on two pairs of parentheses $D_2$) is in order $exp(\Theta(n^2/\ln n))$ \cite{CFRat}, and, hence, this is the upper bound on the value of the rational index for every context-free language.

While both problems is not parallelizable in general, it is useful to develop more efficient parallel solutions for specific subclasses of the context-free languages. For example, there are context-free languages which admit more efficient parallel algorithms in comparison with the general case of context-free recognition \cite{IBARRA2, IBARRA, Okhotin2014ComplexityOI}.  The same holds for the CFL-reachability problem: there are some examples of context-free languages, for which the CFL-reachability problem lies in NL complexity class (for example, linear and one-counter languages) \cite{labelledGraphs, LReach, Regularrealizability, VyalyiRR}. These languages have the polynomial rational index.

The family of linear languages (linear Datalog programs, respectively)
is the well-known subclass of context-free languages
that has polynomial rational index \cite{RatBasic, Ullman}.
The value of its rational index is in $O(n^2)$ \cite{RatBasic}.
Linear Datalog programs have been widely studied by the deductive database community,
because efficient evaluation methods and optimization techniques
exist for such programs \cite{linearisability, linopt, Ullman}.
For instance, the Datalog program in Example~\ref{DatalogExample} is linear.
Many efforts have been devoted to find larger subclasses of context-free languages (Datalog programs)
having the polynomial rational indices
~\cite{ChainQ, linearisability, RatBasic, KanellakisParallel, Regularrealizability, Ullman, VyalyiRR}.
Two equivalent classes generalizing linear languages were proposed:
piecewise linear programs \cite{KanellakisParallel, Ullman}
and the family of quasi-rational languages
(the substitution closure of the family of linear languages) \cite{RatBasic};
both were independently shown to have the polynomial rational index.
Quasi-rational languages are generated by \textit{nonexpansive} grammars.
A variable $S$ in a context-free grammar $G$ is expansive
if there exists a derivation $S  \stackrel {*}{\Rightarrow } uSvSw$ for some words $u, v, w$.
A grammar which contains no expansive variable is said to be nonexpansive.
However, it was shown by Afrati et al.~\cite{linearisability}
that piecewise linear programs is equivalent to linear programs.
Therefore, a generalization of linear languages preserving polynomiality of the rational index
remains to be found.

In this work we investigate the rational index of bounded-oscillation languages.
Bounded-oscillation languages were introduced by Ganty and Valput \cite{BoundOsc}.
Just like the class of linear languages, this class is defined by restriction on the pushdown automata. This restriction is based on the notion of \textit{oscillation}, a special measure of how the stack height varies over time. This class generalizes the family of linear languages.

\textbf{Our contributions.} Our results can be summarized as follows:
\begin{itemize}
\item We show that the rational index of bounded-oscillation languages of Ganty and Valput \cite{BoundOsc} is polynomial and give an upper bound on its value in dependence of the value of oscillation.
\item We give upper bounds on the value of rational indices of previously studied subclasses of bounded-oscillation languages: superlinear and ultralinear languages.
\end{itemize}

\section{Preliminaries}
\label{sec:prel}
\label{preliminaries}
\paragraph{Formal languages.} 
A \textit{context-free grammar} is a 4-tuple $G = (\Sigma, N, P, S)$, where $\Sigma$ is a finite set of alphabet symbols,  $N$ is a set of nonterminal symbols, $P$ is a set of production rules and $S$ is a start nonterinal. $L(G)$ is a context-free language generated by context-free grammar $G$. We use the notation $A \stackrel {*}{\Rightarrow } w$  to denote that the string $w \in \Sigma^*$ can be derived from a nonterminal $A$ by sequence of applying the production rules from $P$. A \textit{parse tree} is an entity which represents the structure of the derivation of a terminal string from some nonterminal.

A grammar $G$ is said to be is in the \textit{Chomsky normal form}, if all production rules of $P$ are of the form:
$A \rightarrow BC$, $A \rightarrow a$ or $S \rightarrow \varepsilon$, where $A, B, C \in N$ and $a \in \Sigma$.

The set of all context-free languages is identical to the set of languages accepted by pushdown automata (PDA). \textit{Pushdown automaton} is a 7-tuple $M = (Q, \Sigma, \Gamma, \delta, q_0, Z, F)$, where $Q$ is a finite set of states, $\Sigma$ is a input alphabet, $\Gamma$ is a finite set which is called the stack alphabet, $\delta$ is a finite subset of $Q \times (\Sigma \cap \{\varepsilon\}) \times \Gamma \times Q \times \Gamma^*$,
$q_{0}\in Q$ is the start state, $Z \in \Gamma$ is the initial stack symbol and
$F\subseteq Q$ is the set of accepting states.

A \textit{regular language} is a language that can be expressed with a regular expression or a deterministic or non-deterministic finite automata.
A \textit{nondeterministic finite automaton} (NFA) is a 5-tuple, $(Q,\Sigma ,\delta ,Q_{0},F)$, where $Q$ is a finite set of states, $\Sigma$ is a finite set of input symbols, $\delta:Q\times \Sigma \rightarrow 2^{|Q|}$ is a transition function, $Q_0 \subseteq Q$ is a set of initial states, $F \subseteq Q$ is a set of accepting (final) states. \textit{Deterministic finite automaton} is a NFA with the following restrictions: each of its transitions is uniquely determined by its source state and input symbol, and reading an input symbol is required for each state transition.

 For a language $L$ over an alphabet $\Sigma$, its rational index $\rho_L$ is a function defined as follows:
$$\rho_L(n) = \max\{\min\{|w|:w \in L \cap K\}, K \in {Rat}_n, L \cap K \neq \emptyset\},$$ where $|w|$ is the length of a word $w$ and ${Rat}_n$ denotes the set of regular languages on an alphabet $\Sigma$, recognized by a finite nondeterministic automation with at most $n$ states.
\paragraph{Bounded-oscillation languages.} 
Oscillation is defined using a hierarchy of \textit{harmonics}. Let $\bar{a}$ be a \textit{push}-move and $a$ be a \textit{pop}-move. Then a PDA run $r$ can be described by a well-nested sequence $\alpha(r)$ of $\bar{a}$-s and $a$-s. Two positions $i<j$ form a \textit{matching pair} if the corresponding $\bar{a}$ at $i$-th position of the sequence matches with $a$ at $j$-th position. For example, word $\bar{a}\bar{a}\bar{a}aa\bar{a}aa$ has the following set of matching pairs: $\{(1, 8), (2, 5), (3, 4), (6, 7)\}$ ($\bar{a}(\bar{a}(\bar{a}a)a)(\bar{a}a)a$).

Harmonics are inductively defined as follows:
\begin{itemize}
\item  order 0 harmonic $h_0$ is $\varepsilon$
\item  $h_{(i+1)}$ harmonic is $\bar{a}h_ia\ \bar{a}h_ia$.
\end{itemize}
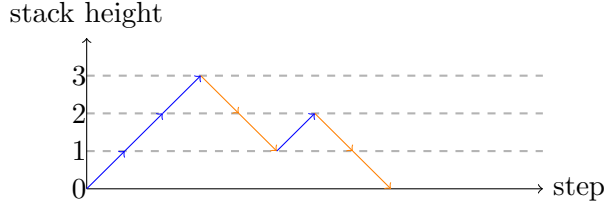
\begin{figure}
\centering
\begin{tikzpicture}
    \draw[thick, dashed, opacity=0.3] (0,0.5) -- (6,0.5);
     \draw[thick, dashed, opacity=0.3] (0,1) -- (6,1);
      \draw[thick, dashed, opacity=0.3] (0,1.5) -- (6,1.5);
      \draw[->] (0,0) -- (6,0) node[right] {step};
      \draw[->] (0,0) -- (0,2) node[above] {stack height};
     \draw[->, blue] (0,0) -- (0.5,0.5);
      \draw[->, blue] (0.5,0.5) -- (1,1);
      \draw[->, blue] (1,1) -- (1.5,1.5);
       \draw[->, orange] (1.5,1.5) -- (2,1);
    \draw[->, orange] (2,1) -- (2.5,0.5);
    \draw[->, blue] (2.5,0.5) -- (3,1);
    \draw[->, orange] (3,1) -- (3.5,0.5);
 \draw[->, orange] (3.5,0.5) -- (4,0);
\node (null) at (-0.1, 0) {0}; 
\node (one) at (-0.1, 0.5) {1}; 
\node (two) at (-0.1, 1) {2}; 
\node (three) at (-0.1, 1.5) {3}; 
    \end{tikzpicture}
\caption{Stack heights during the run of PDA.}
\label{oscb}
\end{figure}

PDA run $r$ is \textit{k-oscillating} if the harmonic of order $k$ is the greatest harmonic that occurs in $r$ after removing $0$ or more matching pairs. \textit{Bounded-oscillation languages} are languages accepted by pushdown automata with all runs $k$-oscillating. It is important that the problem whether a given CFL is a bounded-oscillation language is undecidable \cite{BoundOsc}.
\begin{example}
Consider Figure \ref{oscb}. It shows how the stack height changes during the run of a PDA. Corresponding well-nested word $\alpha(r)$ is $\bar{a}\bar{a}\bar{a}aa\bar{a}aa$. The greatest harmonic in this word is order 1 harmonic (moves forming harmonic are marked in bold, removed matching pairs are $(1, 8)$ and $(2, 5)$): $\bar{a}\bar{a}\mathbf{\bar{a}a}a\mathbf{\bar{a}a}a$, therefore oscillation of the run $r$ is 1.
\end{example}

The oscillation of a parse tree of a context-free grammar can be defined similarly to the oscillation of a PDA run.
Given a parse tree $t$, the corresponding well-nested word $\alpha(t)$ is defined inductively as follows:
\begin{itemize}
\item if $n$ is the root of $t$ then $\alpha(t) = \bar{a}\alpha(n)$
\item if $n$ is a leaf then $\alpha(n)=a$
\item if $n$ has $k$ children then $\alpha(n) = a\underbrace{\bar{a}...\bar{a}}_\text{$k$ times}\alpha(n_1)...\alpha(n_k)$
\end{itemize}

Moreover, given a PDA run $r$, there exists a corresponding parse tree $t$ with the same well-nested word $\alpha(t)=\alpha(r)$ and vice versa \cite{BoundOsc}.

The oscillation of a parse tree is closely related with its $dimension$. For each node $v$ in a tree $t$, its dimension $dim(v)$ is inductively defined as follows:
\begin{itemize}
\item if $v$ is a leaf, then $dim(v)$ = 0
\item if $v$ is an internal node with $k$ children $v_1, v_2, ..., v_k$ for $k \ge 1$, then 
$$
dim(v) = 
 \begin{cases}
   \max_{i \in \{1...k\}}dim(v_i) &\text{if there is a unique maximum}\\
   \max_{i \in \{1...k\}}dim(v_i)+1 &\text{otherwise}
 \end{cases}
$$
\end{itemize}

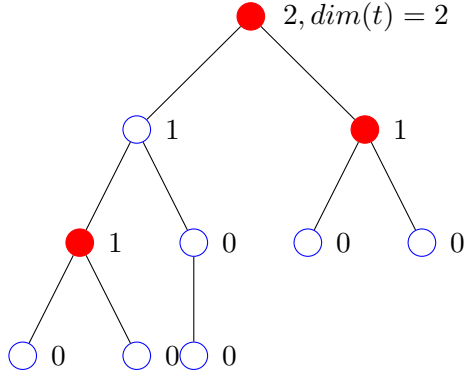
\begin{figure}
\centering
\begin{tikzpicture}[
level 1/.style={sibling distance=3cm},
level 2/.style={sibling distance=1.5cm}]
%\tikzstyle{every node}=[circle,draw]

\node[circle,draw] (Root) [ fill=red, red] {}
    child {
    node[circle,draw, blue] (l) {} 
    child { node[circle,draw, fill=red, red](ll) {}
            child { node[circle,draw, blue] (p) {} }
            child { node[circle,draw, blue] (pl) {} }
             }
    child { node[circle,draw, blue](lr) {} 
          child { node[circle,draw, blue] (plr) {} }
      }
}
child {
    node[circle,draw,  fill=red, red] (r) {}
    child { node[circle,draw, blue] (rl) {}} 
    child { node[circle,draw, blue] (rr) {} }
};
\node  [right=0.05cm of p] {0};
\node  [right=0.1cm of Root] {$2, dim(t)=2$};
\node  [right=0.05cm of l] {1};
\node  [right=0.05cm of r] {1};
\node  [right=0.05cm of ll] {1};
\node  [right=0.05cm of lr] {0};
\node  [right=0.05cm of pl] {0};
\node  [right=0.05cm of plr] {0};
\node  [right=0.05cm of rl] {0};
\node  [right=0.05cm of rr] {0};
\end{tikzpicture}
\caption{A tree $t$ with $dim(t)=2$. Nodes having children without unique maximum are filled.}
\label{oscbtree}            
\end{figure}

The dimension of a parse tree $t$ $dim(t)$ is the dimension of its root. 
It is observable from the definition that the dimension of a tree $t$
is the height of the largest perfect binary tree,
which can be obtained from $t$ by contracting edges and accordingly identifying vertices.
A tree of dimension $dim(t) = 2$ is illustrated in Figure~\ref{oscbtree}.

It is known that the dimension of parse trees and its oscillation are in linear relationship.

\begin{lemma}[\cite{BoundOsc}]
\label{boscdim}
Let a grammar $G = (\Sigma, N, P, S)$ be in Chomsky normal form and let $t$ be a parse tree of $G$. Then $osc(t) - 1 \le dim(t) \le 2osc(t)$.
\end{lemma}
\paragraph{Context-free language reachability.} 
A \textit{directed labeled graph} is a triple $D = (Q, \Sigma, \delta)$, where $Q$ is a finite set of nodes, $\Sigma$ is a finite set of alphabet symbols,
and $\delta \subseteq Q \times \Sigma \times Q$ is a finite set of labeled edges. Let $L(D)$ denote a graph language~--- a regular language, which is recognized by the NFA $(Q,\Sigma ,\delta ,Q, Q)$ obtained from $D$ by setting every state as inial and accepting.

Let $i\pi j$ denote a unique path between nodes $i$ and $j$ of the input graph and $l(\pi)$ denote a unique string obtained by concatenating edge labels along the path $\pi$. Then the CFL-reachability can be defined as follows.
\begin{definition}[Context-free language reachability]
Let $L \subseteq \Sigma^*$ be a context-free language and $D = (Q, \Sigma, \delta)$ be a directed labeled graph. Given two nodes $i$ and $j$ we say that $j$ is \textit{reachable} from $i$ if there exists a path $i \pi j$, such that $l(\pi) \in L$. 
\end{definition}
There are four varieties of CFL-reachability problems: all-pairs problem, single-source problem, single-target problem and single-source/single-target problem \cite{RepsBasic}. In this paper we consider all-pairs problem. The \textit{all-pairs problem} is to determine all pairs of nodes $i$ and $j$ such that $j$ is reachable from $i$.

\section{Rational index of bounded-oscillation languages}
\label{sec:osc}
\subsection{Upper bounds on the rational index of bounded oscillation languages}
Before we consider the value of the rational index for $k$-bounded-oscillation languages, we need to prove the following.
\begin{lemma}
\label{lem:treeheight}
Let  $G = (\Sigma, N, P, S)$ be a context-free grammar in Chomsky normal form,  $D=(V, E, \Sigma)$ be a directed labeled graph with $n$ nodes. Let $w$ be the shortest string in $L(G)\cap L(D)$. Then the height of every parse tree for $w$ in $G$ does not exceed $|N|n^2$.
\end{lemma}

\begin{proof}
Consider grammar $G'$ for $L(G)\cap L(D)$. The grammar $G = (\Sigma, N', P', S')$ can be constructed from $G$ using the classical construction by Bar-Hillel et al.~\cite{BarHillel}: $N' \subseteq N \times V \times V $  contains all tiples $(A, i, j)$ such that $A \in N, i, j \in V$ ; $P'$ contains production rules in one of the following forms:
\begin{enumerate}
\item $(A, i, j) \rightarrow (B, i, k), (C, k, j)$ for all $(i, k, j)$ in $V$  if $A \rightarrow BC \in P$
\item $(A, i, j) \rightarrow a$ for all $(i, j)$ in $V$ if $A \rightarrow a$.
\end{enumerate}
A triple $(A, i, j)$ is \textit{realizable} if and only if there is a path $i\pi j$ such that $A \stackrel {*}{\Rightarrow } l(\pi)$ for some nontermimal $A \in N$. Then the parse tree $t_G$ for $w$ in $G$ can be converted into parse tree $t_{G'}$ in $G'$. Notice that every node of $t_{G'}$ is realizable triple. Also it is easy to see that the height of $t_G$ is equal to the height of $t_{G'}$. Assume that $t_{G'}$ for $w$ has a height of more than $|N|n^2$. Consider a path from the root of the parse tree to a leaf, which has length greater than $|N|n^2$. There are $|N|n^2$ unique labels $(A, i, j)$ for nodes of the parse tree, so according to the pigeonhole principle, this path has at least two nodes with the same label. This means that the parse tree for $w$ contains at least one subtree $t$ with label $(A, i, j)$ at the root, which has a subtree $t'$ with the same label. Then we can change $t$ with $t'$ and get a new string $w'$ which is shorter than $w$, because the grammar is in Chomsky normal form. But $w$ is the shortest, then we have a contradiction.

\end{proof}
From Lemma \ref{lem:treeheight} one can deduce an alternative proof of the fact that the rational index of linear languages is in $O(n^2)$ \cite{RatBasic}: the number of leaves in a parse tree in linear grammar in Chomsky normal form is proportional to its height, and thus it is in $O(n^2)$.
\begin{lemma}
\label{oscbnddim}
Let $G$ be a grammar $G = (\Sigma, N, P, S)$ in Chomsky normal form, such that every parse tree $t$ has $dim(t) \le d$, where $d$ is some constant. Let $D=(V, E, \Sigma)$ be a directed labeled graph with $n$ nodes. Then $\rho_{L(G)}$ is in $O(h^d)$ in the worst case.
\end{lemma}
\begin{proof}
Proof by induction on dimension $dim(t)$.

\textbf{Basis.} $dim(t) = 1$.
\\
Consider a tree $t$ with the dimension $dim(t) = 1$. The root of the tree has the same dimension and has two children (because the grammar is in Chomsky normal form). There are two cases:  first, when both of child nodes have dimension equal to 0, then the tree has only two leaves, and second, when one of the children has dimension 1, and the second child has dimension 0. For the second case we can recursively construct a tree with the maximum number of leaves in the following way. Every internal node of such a tree has two children, one of which has dimension equal to 0 and therefore has only one leaf. This means that the number of leaves (and, hence, $\rho_{L(G)}$) in such a tree is bounded by its height and is in $O(h)$. 
\\
\textbf{Inductive step.} $dim(t) = d + 1$.
\\
Assume that $\rho_{L(G)}$ is at most $O(h^{d})$ for every $d$ in the worst case, where $h$ is the height of the tree. We have two cases for the root node with dimension equal to $d+1$: 1) both of children have a dimension equal to $d$, then by proposition the tree of height $h$ has no more than $O(h^{d})$ leaves; 2) one of the children has a dimension $d + 1$, and the second child $v$ has a dimension $dim(v) \le d$. Again, a tree with the maximum number of leaves can be constructed recursively:  each node of such tree has two children $u$ and $v$ with dimensions $d+1$ and $d$ respectively (the greater the dimension of the node, the more leaves are in the corresponding tree in the worst case). By the induction assumption there are no more than $(h-1)^d + (h-2)^d + (h-3)^d + ... + 1 = O(h^{d+1})$ leaves, so the claim holds for $dim = d+1$.
\end{proof}
Combining Lemma \ref{lem:treeheight} and Lemma \ref{oscbnddim}, we can deduce the following.
\begin{corollary}
\label{finaldim}
Let $G$ be a grammar $G = (\Sigma, N, P, S)$ in Chomsky normal form, such that every parse tree $t$ has $dim(t) \le d$, where $d$ is some constant. Let $D=(V, E, \Sigma)$ be a directed labeled graph with $n$ nodes. Then $\rho_{L(G)}$ is in $O({(|N|n^2)}^d)$ in the worst case.
\end{corollary}
\begin{theorem}
\label{oscbndosc}
Let $L$ be a $k$-bounded-oscillation language with grammar $G = (\Sigma, N, P, S)$ in Chomsky normal form and $D=(V, E, \Sigma)$ be a directed labeled graph with $n$ nodes. Then $\rho_{L(G)}$ is in $O({|N|}^{2k}n^{4k})$ in the worst case.
\end{theorem}
\begin{proof}
By Lemma~\ref{boscdim}, every parse tree of bounded-oscillation language has also bounded dimension. Then the maximum value of the dimension of every parse tree of $k$-bounded-oscillation language is $2k$. By Corollary~\ref{finaldim}, $\rho_{L(G)}$ is in $O({(|N|n^2)}^d)$ and, thus, $\rho_{L(G)}$ does not exceed $O({(|N|n^2)}^{2k}) = O({|N|}^{2k}n^{4k})$.
\end{proof}

As we can see from the proof of Lemma~\ref{oscbnddim}, the family of linear languages is included in the family of bounded-oscillation languages. The reason is that the family of bounded-oscillation languages generalizes the family of languages accepted by finite-turn pushdown automata \cite{BoundOsc}. It is interesting that for general PDA, particularly for $D_2$, the value of oscillation is not constant-bounded: it depends on the length of input and does not exceed $O(\log n)$ for the input of length $n$ \cite*{Gundermann, Wechsung}. However, for some previously studied subclasses of context-free languages,  oscillation is bounded by a constant.

\begin{subsection}{The rational indices of some subclasses of bounded-oscillation languages} 

\paragraph{Superlinear languages.} 
A context-free grammar $G = (\Sigma, N, P, S)$ is \textit{superlinear} \cite{superlinear} if all productions of $P$ satisfy these conditions:
\begin{enumerate}
\item there is a subset $N_L \subseteq N$ such that every $A \in N_L$ has only linear productions $A\rightarrow aB$ or $A\rightarrow Ba$, where $B \in N_L$ and $a \in \Sigma$.
\item if $A \in N \setminus N_L$, then $A$ can have non-linear productions of the form $A \rightarrow BC$ where $B\in N_L$ and $C \in N$, or linear productions of the form $A\rightarrow \alpha B$ $\vert$ $B \alpha$ $\vert$ $\alpha$ for $B \in N_L$, $\alpha \in \Sigma^*$.
\end{enumerate}
A language is \textit{superlinear} if it is generated by some superlinear grammar. 
\begin{theorem}
Let $G$ be a superlinear grammar. Then $\rho_{L(G)}$ is in $O(n^4)$.
\end{theorem}
\begin{proof}
The definition of superlinear grammar implies that
its parse trees have dimension at most 2.
From Corollary~\ref{finaldim}, if dimensions of all parse trees are bounded by some $k$,
then the rational index $\rho_{L(G)}$ of this language is in $O(n^4)$.
\end{proof}
\end{subsection}
\paragraph{Ultralinear languages.} A context-free grammar $G = (\Sigma, N, P, S)$ is \textit{ultrealinear} if there exists a partition $\{N_0, N_1, ..., N_k\}$ of $N$ such that $S \in N_k$ and if $A \in N_i$, where $0 \le i \le k$, then $(A \rightarrow w) \in P$ implies $w \in \Sigma^*N_i\Sigma^*$ or $w \in {(\Sigma \cup N_0 \cup ... \cup N_i-1)}^*$. Such a partition is called an \textit{ultralinear decomposition}. A language is \textit{ultralinear} if it is generated by some ultralinear grammar.

The ultralinear languages were originally defined by Ginsburg and Spanier \cite{Ginsburg1966FiniteTurnPA} as languages recognizable by finite-turn pushdown PDAs (a finite-turn PDA is a PDA with a fixed constant bound on the number of switches between push and pop operations in accepting computation paths). 

Every ultralinear language is generated by an ultralinear grammar in \textit{reduced form} \cite{WORKMAN1976188}.
\begin{definition}[The reduced form of ultralinear grammar.]
An ultralinear grammar $G = (\Sigma, N, P, S)$ is in \textit{reduced form} if its ultralinear decomposition $\{N_0, N_1, ..., N_k\}$ is in the following form:
\begin{enumerate}
\item $N_k=\{S\}$ and $S$ does not appear in the right part of any production rule
\item if $(A \rightarrow w) \in P \setminus \{S \rightarrow \varepsilon\}$ and $A \in N_i$, $0 \le i \le k$, then $w \in (\Sigma \cup N_i\Sigma \cup \Sigma N_i \cup N_jN_j')$, where $j, j' < i$.
\end{enumerate}
\end{definition}
\begin {theorem}
Let $G = (\Sigma, N, P, S)$ be an ultralinear grammar with the ultralinear decomposition $\{N_0, N_1, ..., N_k\}$. Then $\rho_{L(G)}$ is in $O(n^{2k})$.
\end{theorem}
\begin{proof}
Recall that by definition dimension of a parse tree is the height of its largest perfect subtree. Consider the maximum possible size of a perfect subtree which occurs in the parse tree in ultralinear grammar in reduced form. It is easy to see that the rules of the form $A \rightarrow BC$, where $A \in N_i, B, C \in N_{i-1}$ should be used as often as possible to construct the largest binary subtree. Therefore, if grammar has the subset of rules of the form $\{S \rightarrow AB, A \rightarrow A_1A_2, B \rightarrow B_1B_2, A_1\rightarrow A_3A_4, ..., A_i \rightarrow A_{i+2}, A_{i+3}, ...\}$, where $A, B \in N_{k-1}, A_1, A_2, B_1, B_2 \in N_{k-2}, A_3, A_4, ... \in N_{k-3}, ... , A_{i+2}, A_{i+3}, ... \in N_0$, the perfect binary subtree obtained with these rules will be of height not greater than $k$, so the maximum dimension of the parse tree in a ultralinear grammar in reduced form is $k$. By Corollary~\ref{finaldim}  $\rho_{L(G)}$ is in $O(n^{2k})$.
\end{proof}

\section{Conclusions and open problems}
\label{sec:conc}

We have proved that bounded-oscillation languages
have polynomial rational index.
This means that the CFL-reachability problem and Datalog query evalution for these languages is in NC.
This class is a natural generalization of linear languages,
and might be the largest class of queries among such generalizations
that is known to be in NC.
It is interesting whether Datalog programs corresponding to these languages are linearizable
(that is, can always be transformed into linear Datalog programs).

There is a family of languages which has polynomial rational index,
but is incomparable with the linear languages:
\emph{the one-counter languages}.
Moreover, it is not comparable with the bounded-oscillation languages:
for example, the Dyck language $D_1$ is a one-counter language,
but not a bounded-oscillation language for any $k$.
Could this class be generalized in the same manner as linear languages
with respect to the polynomiality of the rational index?
One can consider the Polynomial Stack Lemma by Afrati et al.~\cite{ChainQ},
where some restriction on the PDA stack contents are given,
or investigate the properties of the substitution closure of the one-counter languages,
which is known to have polynomial rational index \cite{RatBasic}.

%%For cite command type as \cite{1}; \cite{3,6} and \cite{2,4,6}.
%%For refcite command type as Refs.~[\refcite{1}];
%%[\refcite{1},\refcite{3}] and [\refcite{1}--\refcite{4}].

\begin{comment}
\section*{Acknowledgments}
This research was supported by the Russian Science Foundation, grant \textnumero 18-11-00100.
\end{comment}

\bibliographystyle{plain}
\bibliography{bounded_oscillation_index}
\end{document}